\documentclass[12pt]{article}
\usepackage{geometry}
\usepackage{amsmath}
\usepackage{amssymb}
\usepackage{bbold}
\usepackage{xspace}
\usepackage{graphicx}

\def\appendix#1{
  \addtocounter{section}{1}
  \setcounter{equation}{0}
  \renewcommand{\thesection}{\Alph{section}}
 \section*{Appendix \thesection\protect\indent \parbox[t]{11.715cm} {#1}}
  \addcontentsline{toc}{section}{Appendix \thesection\ \ \ #1}
  }

\renewcommand{\thefootnote}{\fnsymbol{footnote}}

\numberwithin{equation}{section}

\newcommand{\be}{\begin{equation}}
\newcommand{\ee}{\end{equation}}
\newcommand{\ba}{\begin{aligned}}
\newcommand{\ea}{\end{aligned}}

\newcommand{\ie}{{\it i.e.}}
\newcommand{\eg}{{\it e.g.}}

\newcommand{\sltwo}{\mathfrak{sl}(2)}
\newcommand{\sutwo}{\mathfrak{su}(2)}

\makeatletter
\def\sla@#1#2#3#4#5{{%
  \setbox\z@\hbox{$\m@th#4#5$}%
  \setbox\tw@\hbox{$\m@th#4#1$}%
  \dimen4\wd\ifdim\wd\z@<\wd\tw@\tw@\else\z@\fi
  \dimen@\ht\tw@
  \advance\dimen@-\dp\tw@
  \advance\dimen@-\ht\z@
  \advance\dimen@\dp\z@
  \divide\dimen@\tw@
  \advance\dimen@-#3\ht\tw@
  \advance\dimen@-#3\dp\tw@
  \dimen@ii#2\wd\z@
  \raise-\dimen@\hbox to\dimen4{%
    \hss\kern\dimen@ii\box\tw@\kern-\dimen@ii\hss}%
  \llap{\hbox to\dimen4{\hss\box\z@\hss}}}}
\def\slashed#1{%
  \expandafter\ifx\csname sla@\string#1\endcsname\relax
    {\mathpalette{\sla@/00}{#1}}%
  \else
    \csname sla@\string#1\endcsname
  \fi}
\makeatother

\begin{document}


\thispagestyle{empty}
\begin{flushright}\footnotesize
\texttt{hep-th/0509096}\\
\texttt{AEI-2005-145}\\
\texttt{DESY-05-163}\\
\texttt{ZMP-HH/05-17}\\
\vspace{0.8cm}
\end{flushright}

\renewcommand{\thefootnote}{\fnsymbol{footnote}}
\setcounter{footnote}{0}

\begin{center}
{\Large\textbf{\mathversion{bold}
Stringy sums and corrections to the \\
quantum string Bethe ansatz
}\par}

\vspace{1.5cm}

\textrm{Sakura Sch\"afer-Nameki$^{\alpha}$ and Marija Zamaklar$^{\beta}$ } \vspace{8mm}

\textit{$^{\alpha}$ II. Institut f\"ur Theoretische Physik der Universit\"at Hamburg\\
Luruper Chaussee 149, 22761 Hamburg, Germany} \\
\texttt{sakura.schafer-nameki@desy.de}
\vspace{3mm}

\textit{$^{\alpha}$ Zentrum f\"ur Mathematische Physik, Universit\"at Hamburg\\
Bundesstrasse 55, 20146 Hamburg, Germany} \vspace{3mm}

\textit{$^{\beta}$ Max-Planck-Institut f\"ur Gravitationsphysik, AEI\\
Am M\"uhlenberg 1, 14476 Golm, Germany}\\
\texttt{marzam@aei.mpg.de}
\vspace{3mm}


\par\vspace{1cm}

\textbf{Abstract}\vspace{5mm}

\end{center}

\noindent
We analyze the effects of zeta-function regularization on the
evaluation of quantum corrections to spinning strings. 
Previously, this method was applied in the $\sltwo$ subsector and yielded 
agreement to third order in perturbation
theory with the quantum string Bethe ansatz. 
In this note we discuss
related sums and compare zeta-function regularization against
exact evaluation of the sums, thereby showing that the zeta-function
regularized 
expression misses out perturbative as well as non-perturbative
terms. In particular, this may imply corrections to the
proposed quantum string Bethe equations.  
This also explains the previously observed discrepancy between the
semi-classical string and the quantum string Bethe ansatz in the
regime of large winding number.

\vspace*{\fill}

\newpage
\setcounter{page}{1}
\renewcommand{\thefootnote}{\arabic{footnote}}
\setcounter{footnote}{0}


\tableofcontents


\section{Introduction and Summary}

Explicit checks of the AdS/CFT correspondence beyond the supergravity
approximation have been obstructed by the disjointness of the regimes
in which gauge theory and string theory are understood in perturbation
theory. Exact quantization of string theory on $AdS_5\times S^5$ may help
overcoming this problem and has therefore been the focus of much
recent investigations.   
Key progress in this direction was triggered by the insight
gained from studying the AdS/CFT correspondence in
specific limits, as initiated by \cite{Berenstein:2002jq},
\cite{Gubser:2002tv}, and in \cite{FrolovTseytlinI,  Frolov:2003qc,Frolov:2003tu,Frolov:2003xy,Arutyunov:2003uj,Arutyunov:2003za, Frolov:2004bh}.

Further insight was obtained by identifying the integrable
structures both in gauge and string theory. 
On the gauge theory side, this was deduced from the identification of the planar 
one-loop dilatation operator of $N=4$ SYM 
with the Hamiltonian of an integrable (super) spin chain
\cite{Minahan:2002ve, Beisert:2003yb}, solvable by means of
a Bethe ansatz. The extension of the integrable structure to higher
loops was subsequently shown in \cite{Beisert:2003tq, Beisert:2003ys,
  Serban:2004jf}\footnote{Altough integrability breaks down beyond the
  planar limit, some remnants of it persist and can be used to study
  decays of semi-classical strings 
\cite{Peeters:2004pt}.}. On the other
hand, integrability of the string sigma model on $AdS_5 \times S^5$
\cite{Metsaev:1998it} was observed in
\cite{Bena:2003wd}, and then utilised to test the AdS/CFT
correspondence \cite{Beisert:2003xu, Beisert:2003ea, Engquist:2003rn}\footnote{For reviews and further references see
  \cite{Tseytlin:2003ii, Beisert:2004ry, Zarembo:2004hp, Tseytlin:2004xa,
    Plefka:2005bk}. }. 
An important step linking the two integrable structures on more general
grounds was made in \cite{Kazakov:2004qf} by the
construction of a set of Bethe equations for the classical string
sigma-model\footnote{See also \cite{Arutyunov:2003rg, Arutyunov:2004xy,
  Arutyunov:2005nk, Mikhailov:2005wn} which identified the infinite
  tower of conserved charges on both sides.
The classical string sigma-model reduces in the
  large spin limit to the effective action of the spin-chain, as was
  first observed in \cite{Kruczenski:2003gt}. 
 }.  
These were then compared to the gauge theory Bethe equations in the
thermodynamic limit, first for various subsectors and then the full
$N=4$ SYM and $AdS_5\times S^5$ superstring \cite{Kazakov:2004qf,
  Beisert:2004ag, Schafer-Nameki:2004ik, Beisert:2005bm, Alday:2005gi, 
  Beisert:2005di}. 

Inspired by the classical Bethe equations, a proposal was put forward
for the description of quantum strings on $AdS_5\times S^5$ 
\cite{Arutyunov:2004vx,Staudacher:2004tk,Beisert:2005fw}. It was 
conjectured that the string spectrum
can be described by a new type of quantum string Bethe equations, which
diagonalize some underlying string chain, and which are obtained by
discretizing the classical string Bethe equations
\cite{Kazakov:2004qf}. 
The conjectured quantum string Bethe equations were rigorously tested at infinite
$\lambda $. However, they could potentially receive $1/\sqrt{\lambda
}$ corrections \cite{Arutyunov:2004vx}.

To further test the proposal of
\cite{Arutyunov:2004vx,Staudacher:2004tk,Beisert:2005fw}, 
a detailed comparison between the one-loop worldsheet
correction to the energy of a particular string configuration (which
was computed
semi-classically) to the finite size
corrections following from the quantum
string Bethe ansatz was recently performed \cite{Schafer-Nameki:2005tn}.  
The configuration studied was a circular string
spinning in $AdS_3\times S^1$ \cite{Park:2005ji}. In this case the
correction to the classical energy depends on two
parameters $\mathcal{J}$ and $k$ ($\mathcal{J}^2 =1/\lambda' =
J^2/\lambda$), where $k$ is the 
string winding number and $J$ is the spin in the $S^1$ direction. In 
\cite{Schafer-Nameki:2005tn} the comparison between
semi-classical strings and Bethe ansatz was studied in the following two regimes: 
large $\mathcal{J}$ (and finite $k$) and large
$k$ (and finite $\mathcal{J}$).

In the first instance, due to the high complexity of the sums for the
semi-classical string corrections, the analysis was performed by first
expanding the summands in the parameter $1/{\mathcal J}$ (assuming
that the summation index $n$ is smaller than $\mathcal{J}$) and
subsequent resummation. This procedure clearly breaks down for
$n\geq\mathcal{J}$, and thus yields divergent expressions at each
order in $1/\mathcal{J}^{2l}$. However upon zeta-function
regularisation these agree with the Bethe ansatz in the first three
orders in $1/\mathcal{J}^2$ \cite{Schafer-Nameki:2005tn}. This
extended the leading order agreement previously found in
\cite{Beisert:2005mq, Hernandez:2005nf}. Other
discussions of $1/J$ corrections have appeared in
\cite{Beisert:2003xu, Freyhult:2004iq, Freyhult:2005fn, Fuji:2005ry, Minahan:2005ux}. 

In the second case of large winding number $k$, exact evaluation of
the sum  (which did  not involve 
zeta-function regularization) resulted in a disagreement with the
prediction of the
string Bethe ansatz already at leading order in $1/k$
\cite{Schafer-Nameki:2005tn}. A similar mismatch was observed numerically. 

As a possible explanation for the incompatibility of these results
it was proposed that zeta-function regularization may not correctly
sum the semi-classical string result \cite{Schafer-Nameki:2005tn}.  
A numerical analysis was performed to confirm this conjecture, but due to the
insufficient numerical precision it was not possible to deduce a firm
conclusion in its favour.

In this note we further examine this issue. We find strong evidence that
zeta-function regularization does not give the correct answer for the
sums in question.  We first consider a simple toy example of a sum
which has the same divergence problems when expanded in
$1/\mathcal{J}$ as the sum in \cite{Schafer-Nameki:2005tn}. We then
discuss the case of the folded string in the $\sltwo$ subsector and
circular string in the $\sutwo$ subsector
\cite{Frolov:2003qc,Frolov:2003tu}. We evaluate the sums in
question first by zeta-function regularization and then exactly, using
various methods developed in \cite{Lucietti:2003ki, Lucietti:2004wy,
Schafer-Nameki:2005tn}. These results confirm that zeta-function
regularization does not reproduce the full sum.  The explicit analysis
(in the $\sutwo$ subsector) shows that although the coefficients of
$1/\mathcal{J}^{2n}$ in the expansion are correctly reproduced by the
zeta-function regularisation, the coefficients of
$1/\mathcal{J}^{2n+1}$ are not present, as well as the possibly
non-vanishing non-perturbative contributions ({\it i.e.} of order
$e^{-\mathcal{J}}$).  Both types of terms do not follow from the
quantum string Bethe equations, explaining thus the mismatch
in the large $k$ regime found in \cite{Schafer-Nameki:2005tn}. In particular
the oscillatory behaviour observed in the large $k$ limit in
\cite{Schafer-Nameki:2005tn}, is hidden in the exponential terms,
which are entirely missed by zeta-function regularization.\footnote{We are
grateful to K. Zarembo for this remark.}

One important outcome of this analysis is that the terms in the string
sums which are not captured by the quantum Bethe equations are non-analytic
in the coupling, being proportional to $(\sqrt{\lambda'})^{2n+1}$ for
integral $n$ and $e^{- 1/\sqrt{\lambda'}}$. 
It would be important to modify the S-matrix of
\cite{Arutyunov:2004vx,Staudacher:2004tk,Beisert:2005fw} to
incorporate these effects. Some of these issues are discussed in
\cite{BT}, where the terms with odd powers of $1/\mathcal{J}$ were also
found in the $\sutwo$ subsector and the relation to the Bethe
ans\"atze in \cite{Arutyunov:2004vx,Staudacher:2004tk,Beisert:2005fw}
was discussed. 

The plan of this note is as follows. We first discuss two relatively
simple sums (a toy model, as well as the folded string solution), which can be
evaluated both exactly and by zeta-function regularization. 
In both cases zeta-function regularization fails to reproduce the exact sum. 
In section 4 we apply an approximation method, replacing the sum by an
integral. 
Comparison with the exact expression for the sums, shows that the
approximate evaluation correctly reproduces the terms
missing in the zeta-function regularized result.  
We then apply this method to the $\sutwo$ string and by comparing it with the
zeta-function evaluated result, identify the missing terms.


\section{Folded string solution}

In this section we consider the one-loop
 energy shift for the folded rigid string, which rotates with a single
 spin $S$ in $AdS_3$ and no spin in $S^5$.  This correction was
 computed in \cite{FrolovTseytlinI}, and is (in approximation) given by  
\be \label{ERot}
 \kappa \delta E_{fold} = \sum_{n=1}^\infty \sqrt{n^2 + 4 \kappa^2} +2
 \sqrt{n^2 + 2 \kappa^2}+ 5 n - 8 \sqrt{n^2 + \kappa^2} \, , 
\ee 
where $\kappa \sim \log \mathcal{S}, \, \mathcal{S} = S/\sqrt{\lambda}$.
We wish to evalute this sum for large values of the parameter
 $\kappa$.\footnote{We thank A.~Tseytlin for the suggestion to consider
 this sum.} Recall, that the asymptotic value for the sum, obtained in
 \cite{FrolovTseytlinI} by replacing the sum with an integral is
 \be
\label{FTAsymp}
 \delta E_{fold}^{FT} = - 3 \log 2 \ \kappa+
 O(\kappa^0) \,.  
\ee 
In the following sections we shall evaluate the sum (\ref{ERot}) first
by naive zeta-function regularization and then by various exact
evaluation methods. This will show that zeta-function fails to
reproduce the correct sum.


\subsection{Zeta-function regularization}

Let us first evaluate the sum along the lines of the zeta-function
regularization applied
in \cite{Schafer-Nameki:2005tn}. In order to do so, we pull the
large-$\kappa$ limit into the sum, {\it i.e.} expand each summand in
$1/\kappa$ assuming that the summation index $n$ is smaller than
$\kappa$. This expansion is obviously incorrect when $n \geq
\mathcal{\kappa}$, which reflects itself in the divergence of the resulting sums at each order
in $1/\kappa$ -- despite the fact that the initial sum is convergent.
We regularize these divergences using the zeta-function $\zeta (z)$
analytically continued to negative integers. This can in fact be done
to all orders in $1/\kappa$ and results in
\be \ba
\label{zeta-3}
\delta E_{fold}
&= \sum_n 2 (\sqrt{2}-3) + {1\over \kappa} \sum_n 5n +
O\left({1\over \kappa^2}\right) \cr
&= (3-\sqrt{2}) - {5\over 12} {1\over \kappa}  + O(e^{-\kappa})\, .
\ea
\ee
Here we used that $\zeta (-1) = -B_2/2 = -1/12$ and each higher term
is a sum over $n^{2l}$, and thus vanishes in the zeta-function
prescription. This clearly contradicts the asymptotics in
(\ref{FTAsymp}) by missing out the crucial linear term in
$\kappa$. The result (\ref{FTAsymp}) was obtained by an
approximative method, so it would be desirable to have independent
checks of the sum to confirm the failure of zeta-function
regularization. We shall subsequently present three methods which will
be in agreement with (\ref{FTAsymp}), as well as produce  subleading
terms obtained in (\ref{zeta-3}) (up to exponentially small corrections).


\subsection{Asymptotic evaluation}

A method to asymptotically 
evaluate sums of the type (\ref{ERot}) was obtained in appendix B of
\cite{Lucietti:2003ki} in the context of plane-wave string field
theory. The main idea is to represent the square root
terms using the integral representation of the Gamma-function
\be\label{GammaInt}
{1\over x^z} = {1\over \Gamma(z)} \int_0^\infty dt t^{z-1} e^{-xt} \,,
\ee
which is valid for $x, z>0$. For this to be applicable, we first act with 
${\partial\over \partial \kappa }\left({1\over \kappa}
{\partial\over \partial \kappa} \right )$
 on the sum (\ref{ERot}), which reduces to the expression 
\be
R= -8\kappa \sum_{n=1}^\infty \bigg( {2 \over (n^2 + 4 \kappa^2)^{3/2}}
                             +{1 \over (n^2 + 2 \kappa^2)^{3/2}}
                             -{1 \over (n^2 + \kappa^2)^{3/2}} \bigg) \,.
\ee
Each partial sum is now absolutely convergent and can be asymptotically evaluated
separately using (\ref{GammaInt}). The relevant asymptotics derived in
\cite{Lucietti:2003ki}\footnote{Similar sums are discussed in
  \cite{Foerger:1998kw, Bigazzi:2003jk, Bertoldi:2004rn}.} are
\be\label{LSSAsymp}
\ba
\sum_{n=1}^\infty {1\over (\mathcal{J}^2 + n^2)^{3/2}}
&= {2\over \sqrt{\pi} \mathcal{J}^3} \int_0^\infty ds s^{1/2} e^{-s}
      (\theta (s/(\pi \mathcal{J}^2)) -1)\cr
&= {1\over \mathcal{J}^2} -{1\over 2 \mathcal{J}^3} +
           O\left(e^{-\mathcal{J}}\right)\,.
\ea
\ee
Here $\theta (t)= \sum_{n\in\mathbb{Z}} e^{-\pi  n^2 t}$ and we modular
transformed and used the asymptotics $\theta(t)\rightarrow 1$ as
$t\rightarrow \infty$. Applied to the present case we obtain 
\be
R ={1\over \kappa^2} (3-\sqrt{2}) + O(e^{-\kappa})  \,,
\ee
which after repeated integration results in
\be\label{RotAsymp}
\delta E_{fold} = (3-\sqrt{2}) + {c_1 \kappa \over 2} + {c_0\over
  \kappa} + O(e^{-\kappa})\,,
\ee
where $c_i$ are integration constants, which need to be determined in
some other way. In particular, this is in accord with
\cite{FrolovTseytlinI}, as there are choices for $c_i$, for which the
sums can be made to agree. The integration constants can be derived in
the way done in \cite{Lucietti:2004wy}, but we shall present two
alternative methods to compute the sum exactly.


\subsection{Bessel function evaluation}

The energy shift can be likewise evaluated using the following
integral representation obtained in \cite{Schafer-Nameki:2005tn}
eq. (2.7) and (2.10). Recall that
\be
\sum_{n=1}^\infty 
\left(
\sqrt{(n+ \gamma)^2 + \alpha^2} + \sqrt{(n-\gamma)^2 + \alpha^2} - 2n
-{\alpha^2\over n}\right)
= \gamma^2 - \sqrt{\gamma^2 +\alpha^2} + F(\{\gamma\}, \alpha) \,,
\ee
where we defined the function
\begin{equation}\label{FDef}
 F(\beta ,\alpha )\equiv
 \sqrt{\alpha ^2+\beta ^2}-\beta ^2+\alpha ^2\int_{0}^{\infty }
 \frac{d\xi }{e^\xi -1}\left(
 \frac{2J_1(\alpha \xi )}{\alpha \xi }\,\cosh \beta \xi -1
 \right).
\end{equation}
For large $\alpha$ the asymptotic behaviour of this function is
\begin{equation}\label{FAsymp}
 F(\beta ,\alpha )=-\alpha ^2\ln\left(\frac{e^{C-1/2}}{2}\,
 \alpha \right)+\frac{1}{6}+
 O\left(e^{-\alpha }\right) \, ,
\end{equation}
where $C=0.5772\ldots $ is the Euler constant. 
Applying this to (\ref{ERot}) results in
\begin{equation}\label{sum-asy}
 \delta E_{fold}=-3\ln 2\,\kappa +3-\sqrt{2}-\frac{5}{12\kappa }
 +O(e^{-\kappa }),
\end{equation}
in agreement with \cite{FrolovTseytlinI} and implying that the
integration constants in (\ref{RotAsymp}) are $c_0=-5/12$ and $c_1= -6
\log (2)$. Note that this also calculates all
subleading terms up to exponential (powerlike in $1/\mathcal{S}$ as
$\kappa \sim \log \mathcal{S}$)
corrections.


\subsection{Generalized zeta-function evaluation}

The result obtained with Bessel functions in the last subsection can
be confirmed by the following analytic continuation argument.
Consider a generalization of the Riemann zeta-function
\be
\zeta(s,\kappa) = \sum_{n=1 }^\infty {1\over (n^2+ \kappa^2)^s} \,.
\ee
This is to begin with not well-defined for the choice $s=-1/2$ that we
are interested in, but the generalized zeta function can be
analytically continued to this value. Again, representing the summand
using the Gamma function integral representation as (\ref{LSSAsymp})
derived in appendix B of \cite{Lucietti:2003ki}, 
it follows that the large $\kappa$ asymptotics of this expression is 
\be\label{GeneralizedZeta}
\zeta (s, \kappa) = -{1\over 2} \kappa^{2 s} + 
                   {1\over 2 \kappa^{2s-1}} {\Gamma(1/2) \Gamma (s-1/2)
                    \over \Gamma (s)}  + O(e^{-\kappa}) \,.
\ee
Note now, that this would have been obtained likewise by approximating
the sum by an integral, namely setting $u = n/\kappa$ in the large
$\kappa$ limit
\be\label{IntegralRep}
\zeta (s, \kappa) 
    \sim {1\over \kappa^{2s-1}}\int_0^\infty du {1\over (1+ u^2)^s} 
    =    {1\over 2 \kappa^{2s-1} } {\Gamma (1/2) \Gamma (s-1/2) \over
      \Gamma (s)} \,. 
\ee
Applying this to $\delta E_{fold}$ for $s=-1/2 + \alpha$ for $\alpha
\rightarrow 0$ and that the Riemann zeta-function analytically
continued gives $\zeta
(-1)=-1/12$, we arrive at
\be
\delta E_{fold} = -3 \log2 \ \kappa + (3-\sqrt{2}) -{5\over 12\kappa} + O(e^{-\kappa})  \,,
\ee
in agreement with the above Bessel function evaluation and
\cite{FrolovTseytlinI}. 

This method is quite general and also explains why zeta-function
regularization does not always work. Namely, zeta-function
regularization drops the term that comes from the Gamma-functions in
(\ref{GeneralizedZeta}). 


\subsection{Exponential corrections}

So far we have refrained from working out explictly the exponential
corrections at $O(e^{-\kappa})$. These may however turn out to be
crucial for comparison to the quantum string Bethe ansatz. We shall now prove that
in the simpler case of the folded string these terms are indeed
non-vanishing and find explicit formulas for these terms. 
As the starting point, consider the asymptotic evaluation method presented
earlier. Recall that
\be
\kappa \sum_{n=1}^\infty {1\over (n^2 + \kappa^2 a^2)^{3/2}} 
= -{1 \over 2 a^3 \kappa^2} + {1\over a^2 \kappa}+ {2 \over a^2
  \kappa} \int_0^\infty dt e^{-t} \sum_{n=1}^\infty \left(e^{-\pi^2 n^2 \kappa^2 a^2
  /t}\right) \,.
\ee
The last term is the exponential correction term and can be further evaluated
\be
\ba
R_{exp} 
&= {2 \over a^2\kappa} \sum_{n=1}^\infty \int_0^\infty dt e^{-t -\pi^2 n^2
     a^2 \kappa^2/t}  \cr
&= {4\over a^2} \sum_{n=1}^\infty 2 \pi a n K_1(2 \pi n a\kappa) \,.
\ea
\ee
Note that $\partial_\kappa K_0(2
\pi n a \kappa) = - 2\pi n a K_1 (2 \pi n a \kappa)$. So, already
integrating up once with respect to $\kappa$ yields
\be
\int d\kappa R_{exp} = -{4 \over a^2} \sum_{n=1}^\infty  K_0(2 \pi n a
\kappa) \,.
\ee
Then apply the integral represetation (see also appendix D of
\cite{Lucietti:2004wy})
\be
K_0(z \kappa ) = \int_0^\infty dt {e^{-z \sqrt{t^2 + \mu^2}} \over
  \sqrt{t^2 + \mu^2}} \,,
\ee
and perform the sum, which yields
\be
\ba
\kappa \int d \kappa R_{exp} 
&=-{4 \over a^2}\kappa  \int_0^\infty dt
 {1\over \sqrt{t^2 + \kappa^2}} {1\over e^{2 \pi a \sqrt{t^2 +
	 \kappa^2}} -1} \cr
&= -{2 \over a^2} \kappa \int_1^{\infty} dr {\coth (a \kappa \pi r )-1
   \over  \sqrt{s^2 -1}} \,.
\ea
\ee
Integrating repeatedly with respect to $\kappa$, we arrive at
\be
\int d\kappa \kappa \int d\kappa R_{exp} 
= -{2 \over a^2}
     \int_1^\infty dr 
     \left[    
         {\kappa \log \left(1-e^{-2 \pi a \kappa r}\right) 
          \over a r
	   \sqrt{r^2 -1}}
        +{(r^3 + r^2 -1) {\rm Li}_2 \left( e^{-2\pi a \kappa r}\right)
	   \over 2 a^2 \pi^2 r^2 \sqrt{r^2 -1}}   
    \right] \,.
\ee
This is a closed formula for the exponential correction term we were looking for. 
Adding up the contributions with the various choices for $a$ of each summand in
(\ref{ERot}) produces the complete correction term for the folded string. 

If one is interested in obtaining the first correction term in
$e^{-\kappa} O(\kappa^0)$ explicitly, one can proceed as follows. 
Note that $\int d\kappa \kappa K_0(b \kappa) = -\kappa K_1(b \kappa)/b$. So
we obtain
\be
\int d\kappa \kappa \int d\kappa R_{exp} = {2 \kappa \over \pi a^3}
\sum_{n=1}^\infty {K_1 (2 \pi n a \kappa) \over n} \,.
\ee
With the asymptotics $K_1 (z) = \sqrt{\pi/2 z } e^{-z} (1+ O(1/z))$ we
obtain that the first exponential correction terms are
\be
\int d\kappa \kappa \int d\kappa R_{exp}
= {\kappa \over \pi a^3}\sum_{n=1}^\infty 
    e^{- 2 \pi n a \kappa} {1\over n} {\sqrt{1 \over n a \kappa }} 
  \left[1+ O\left({1\over \kappa }\right)\right] \,.
\ee
Adding together the terms with the correct prefactors and choices for
$a$ gives the correction to (\ref{ERot}).

In summary we have shown in this section that the exponential
corrections do not vanish for the folded string case. It would of
course be interesting to see, whether they contribute in more
complicated sums than (\ref{ERot}), such as the one-loop energy shift
for the $\sutwo$ and $\sltwo$ subsectors.


\section{Toy model}

As a second test case consider the situation of two bosonic and two
fermionic frequencies with the energy shift given by
\be\label{ToyE}
\delta E_{toy}= \sum_{n=1}^\infty
  \sqrt{1+ (n+\gamma)^2/\mathcal{J}^2} +
  \sqrt{1+ (n-\gamma)^2/\mathcal{J}^2} - 2 \sqrt{1+ n^2/ \mathcal{J}^2} \,,
\ee
where $\gamma$ is a constant independent of $\mathcal{J}$ and the sum
is convergent in the same sense as for the $\sutwo$ and $\sltwo$
spinning strings. 
Again we compare zeta function regularization with the
exact evaluation of the sum in the large $\mathcal{J}$ limit 
and find disagreement. 


\subsection{Zeta-function regularization}

For the naive perturbative evaluation of (\ref{ToyE}), pull the large
$\mathcal{J}$ limit through the sum. 
As each term in the $1/\mathcal{J}$ expansion is of order $n^0$ or
higher, using zeta-function regularization the sum evaluates to
\be
\label{zeta-regu}
\delta E^\zeta_{toy} = -{1\over 2} \left(-2 + 2 \sqrt{1+ {\gamma^2\over
    \mathcal{J}^2}}\right)  \,.
\ee
Expanding this in $1/\mathcal{J}$ yields the energy shift at
arbitrary loop orders as obtained from this prescription.


\subsection{Asymptotic evaluation of sums}

Alternatively, in this simple case, one can evaluate the sum exactly
(up to terms $e^{-\mathcal{J}}$) using the method in \cite{Lucietti:2003ki}.
Consider the sum
\be
\delta E_{toy} \mathcal{J}= S
= \sum_{n=1}^\infty
\sqrt{(n+\gamma)^2 + \mathcal{J}^2 } +
\sqrt{(n-\gamma)^2 + \mathcal{J}^2}  -2
\sqrt{n^2 + \mathcal{J}^2 } \,.
\ee
Then following the strategy in \cite{Lucietti:2003ki}, act with
${\partial\over \partial \mathcal{J}}\left({1\over \mathcal{J}}
{\partial\over \partial \mathcal{J}} \right )$ to obtain
\be
R= -\mathcal{J}
\sum_{n=1}^\infty
  {1 \over ((n+\gamma)^2 + \mathcal{J}^2 )^{3/2}}
+ {1 \over ((n-\gamma)^2 + \mathcal{J}^2)^{3/2}}
-2{1 \over (\mathcal{J}^2 + n^2)^{3/2}} \,.
\ee
Now each part of the sum is absolutely convergent by itself and can be evaluated and
later on integrated up to give the result for the complete sum.
The last summand is easiest and is evaluated the same way as in
appendix B of \cite{Lucietti:2003ki}, {\it i.e.} (\ref{LSSAsymp}).
The remaining two terms are computed likewise. First recall the definition of the
generalized theta-functions
\be
\theta\bigg[\ba a\cr b \ea\bigg] (t) = \sum_{n=-\infty}^\infty e^{\pi t
  (n+a)^2 +  2 \pi  n b i } \,,
\ee
which satisfies the modular transformation law, shown by Poisson resummation,
\be
\theta\bigg[\ba a\cr b \ea\bigg] (t) = {1\over \sqrt{-t}}
\theta\bigg[\ba b\cr -a \ea\bigg] (1/t)\,.
\ee
So in particular we can write
\be
\theta \bigg[\ba \gamma \cr 0\ea \bigg] (-t/\pi) = e^{-\gamma^2 t } +
\sum_{n=1}^\infty \left(e^{-(n+\gamma)^2 t} + e^{-(n-\gamma)^2 t}\right)\,.
\ee
This allows the evaluation of the remaining two terms in the sum,
again asymptotically for large $\mathcal{J}$
\be\label{Comput}
\ba
\sum_{n=1}^\infty
& {1 \over ((n+\gamma)^2 + \mathcal{J}^2 )^{s}}
+ {1 \over ((n-\gamma)^2 + \mathcal{J}^2)^{s}} \cr
& \qquad=
    {1\over \Gamma (s)} \int_0^\infty dr r^{s-1} e^{-\mathcal{J}^2 r}
     \sum_{n=1}^\infty \left(e^{-(n+\gamma)^2 r} + e^{-(n-\gamma)^2
       r}\right)\cr
& \qquad=
    {1\over \Gamma(s)\mathcal{J}^{2s}} \int_0^\infty dt t^{s-1} e^{- t}
     \left(\theta \bigg[\ba \gamma \cr 0\ea \bigg] (-t/(\pi\mathcal{J}^2))-
    e^{-\gamma^2 t/\mathcal{J}^2} \right) \cr
& \qquad= -{1\over( \mathcal{J}^2 + \gamma^2)^{s}} 
     + {\sqrt{\pi}\over \Gamma(s) \mathcal{J}^{2s-1}}
        \int_0^\infty dt  t^{s-3/2}e^{-t} \theta \bigg[\ba 0 \cr -\gamma\ea \bigg]
    (-\pi \mathcal{J}^2/t)\cr
& \qquad= -{1\over( \mathcal{J}^2 + \gamma^2)^{s}} 
          + { \sqrt{\pi} \Gamma (s-1/2) \over \Gamma (s) \mathcal{J}^{2s-1}} 
          + {\sqrt{\pi} \over \mathcal{J}^{2s-1}}
             \int_0^\infty dt t^{s-3/2} 
             e^{-t} \left(\theta \bigg[\ba 0 \cr -\gamma\ea \bigg]
    (-\pi \mathcal{J}^2/t) -1 \right) \,.
\ea
\ee
For $s=3/2$ the last term is of order $e^\mathcal{-J}$, which can be
seen by changing to $u=\mathcal{J}^N t$. 
So in summary we obtain
\be
\sum_{n=1}^\infty
  {1 \over ((n+\gamma)^2 + \mathcal{J}^2 )^{3/2}}
+ {1 \over ((n-\gamma)^2 + \mathcal{J}^2)^{3/2}}
= {2\over \mathcal{J}^2} -{1\over( \mathcal{J}^2 + \gamma^2)^{3/2}}
+ O\left(e^{-\mathcal{J}}\right) \,.
\ee
Thus we obtain that
\be
R= -{\mathcal{J}} \left({1\over \mathcal{J}^3} - {1\over
  (\mathcal{J}^2 + \gamma^2)^{3/2}}\right) \,.
\ee
Integrating up, we obtain
\be
\delta E_{toy} = {1\over \mathcal{J}} \left(\mathcal{J} - \sqrt{\gamma^2 +
  \mathcal{J}^2} \right) + c_0 \mathcal{J}+ {c_1 \over \mathcal{J}}
+ O(e^{-\mathcal{J}}) \,,
\ee
which for vanishing integration constants agrees up to terms $O(e^{-\mathcal{J}})$ with the
perturbative zeta-function regularized expression $\delta E^\zeta$.

In order to determine the integration constants, derive with respect
to $\gamma$ and then evaluate the large ${\mathcal J}$ in analogy to
\cite{Lucietti:2004wy}. However, we shall determine these using the Bessel
and generalized zeta-function methods introduced earlier. 


\subsection{Bessel function evaluation}

Consider now the evaluation using Bessel functions. 
First split the sum into two partial sums which both converge absolutely
\begin{eqnarray}
\delta E_{toy} \mathcal{J}= S 
&=& S_1 + S_2 \nonumber \\
\label{S1}
S_1 &=& \sum_{n=1}^\infty
\sqrt{(n+\gamma)^2 + \mathcal{J}^2 } +
\sqrt{(n-\gamma)^2 + \mathcal{J}^2} - 2n - {\mathcal{J}^2\over n} \nonumber \\
\label{S2}
S_2 &=&  -2  \sqrt{n^2 + \mathcal{J}^2 }+ 2n + {\mathcal{J}^2\over n} \,.
\end{eqnarray}
The representation (\ref{FDef}) implies
\begin{eqnarray}
\label{S12}
S_1 &=& \gamma^2 - \sqrt{\gamma^2 + {\mathcal J}^2 } + F(\{ \gamma \}, {\mathcal J}) \nonumber \\
S_2 &=&  {\mathcal J} - F(0, \mathcal{J})\,.
\end{eqnarray}
The large $\mathcal{J}$ asymptotics follow from (\ref{FAsymp}), so that
\begin{eqnarray}
\label{sums12}
S_1 &=& \gamma^2 - \sqrt{\gamma^2 + \mathcal{J}^2 } - \mathcal{J}^2   \ln \mathcal{J} - \mathcal{J}^2 \ln \left( {e^{C- {1\over 2}} \over 2} \right) + {1\over 6} + O\left(e^{-\mathcal{J} }\right)
 \nonumber \\
\label{S2}
S_2 &=&  \mathcal{J}+ \mathcal{J}^2  \ln  \mathcal{J} + \mathcal{J}^2
\ln \left( {e^{C- {1\over 2}} \over 2} \right) - {1\over 6} +
O\left(e^{-\mathcal{J} }\right) \,,
\end{eqnarray}
and thus the asymptotic expansion for the energy is up to
exponentially small corrections
\be \label{EtoyBessel}
\delta E_{toy} 
= {1\over \mathcal{J}}\left(\gamma^2+\mathcal{J} - \sqrt{\gamma^2 + \mathcal{J}^2}
       \right) + O\left(e^{-\mathcal{J} }\right)  \,, 
\ee
This is in agreement with the asymptotic evaluation and determines the
integration constants as $c_0=0$ and $c_1=\gamma^2$.


\subsection{Generalized zeta-function evaluation}

To confirm the result from the last section, we apply analytic
continuation to the following generalized zeta-function 
\be
\zeta (s, \gamma,\mathcal{J}) = \sum_{n=1}^\infty {1\over ((n+ \gamma)^2 +
  \mathcal{J}^2)^s}\,.
\ee
Then by analytic continuation to $s=-1/2$ we can compute the sums in
$\delta E$. The asymptotics for large values of $\mathcal{J}$ follow
using (\ref{Comput}) in
the last section using generalized theta functions and setting $s=-1/2$
\be
\zeta (s, \gamma,\mathcal{J}) + \zeta (s, -\gamma,\mathcal{J}) = 
 -{1\over (\gamma^2 + \mathcal{J}^2)^{s}} 
+ {\sqrt{\pi} \Gamma (s-1/2)\over \Gamma(s)\mathcal{J}^{2s-1}} +
\gamma^2 + O(e^{-\mathcal{J}})\,.
\ee
The last term in (\ref{Comput}) for $s=-1/2$ is not exponentially
suppressed and is extracted by performing the integral yielding $\sum
{a_n/(n\mathcal{J})^2} K_1(\pi \mathcal{J} n)$, which has the given asymptotics.
Up to exponential corrections we obtain that the sum has large
$\mathcal{J}$ behaviour given by
\be
\ba
S&= \lim_{\alpha\rightarrow 0}\left\{
    \zeta (-1/2 + \alpha, \gamma, \mathcal{J})+ 
     \zeta (-1/2 + \alpha, -\gamma, \mathcal{J})
  - 2 \zeta (-1/2 +\alpha , 0, \mathcal{J}) \right\}\cr
 &= \lim_{\alpha\rightarrow 0}\left\{ 
         -(\gamma^2 + \mathcal{J}^2)^{1/2 -
          \alpha} + {\sqrt{\pi} \Gamma (-1 +\alpha)\over
	   \Gamma(-1/2)}\mathcal{J}^2 + \gamma^2
          - 2 \left(-{1\over 2\mathcal{J}} + {\mathcal{J}^2\over 2}
   {\Gamma(1/2) \Gamma (-1+ \alpha) \over \Gamma(-1/2 + \alpha) }\right) \right\}     \cr
 &= \gamma^2 + \mathcal{J} - \sqrt{\gamma^2 + \mathcal{J}^2} \,.
\ea
\ee
This is again in agreement with the two independent methods of
evaluation presented earlier and confirms the incompleteness of the
evaluation by means of 
zeta-function regularization.

\section{Zeta-function regularization versus exact summation}

In the previous sections we have performed exact, analytic evaluations of
the sums (\ref{ERot}) and (\ref{ToyE}) using several methods. These were
compared to the zeta function regularized expressions
(\ref{zeta-3}) and (\ref{zeta-regu}) and were found to disagree with
them. We would now like to determine the origin of this
disagreement\footnote{Some of the ideas in this section arose in
  discussions with A.~Tseytlin. 
Similar observations have recently appeared in \cite{BT}.}.  
The nature of this section is more experimental and it would be
important to understand this in full generality, {\it e.g.} in
relation with the observation in (\ref{IntegralRep}). In particular,
it should be possible to extend this to the case of the $\sltwo$
subsector.

To proceed, we split the infinite sum into a finite sum, where
zeta-function regularization applies, and another part, which will be
approximated by simply replacing the sum by an integral. The
correction terms that are computed by the Euler-Maclaurin summation
formula will be discussed below. 
More precisely
\begin{eqnarray} 
\label{sp}
S(\eta) &=& \sum_{n=1}^K f( n, \eta) + \sum_{n=K}^\infty f( n, \eta) \nonumber \\
  &=&  S^{\rm I} (K, \eta) + S^{\rm II} (K, \eta) \, , \quad \quad  K \gg 1 \,,
\end{eqnarray}
 where we have denoted the large parameters $\kappa$ and ${\mathcal J}$ in
 (\ref{ERot}) and (\ref{ToyE}) by $\eta$. Since $K\gg
 1$ the second sum $S^{\rm II}(\eta)$ can be replaced with an
 integral, which will be denoted by $\tilde{S}^{\rm II}$.
 Further let us assume that
\begin{equation}
\label{con2}
1 \ll K\ll \eta \,.
\end{equation}
Then the second sum ({\it i.e.} integral) $\tilde{S}^{{\rm II}}(\eta)$ can be
expanded in $1/\eta$. 

On the other hand, for the zeta-function regularization used in
\cite{Schafer-Nameki:2005tn} one first expands $f(n,\eta)$ in
$1/\eta$ and then resums the expanded series. It is clear that this
expansion fails, when $n>\eta$, inducing spurious divergences. These
were cured by introducing the zeta-function regularization, which
effectively means that one multiplies all terms in the sum with a
factor $e^{-\alpha \, n}$.  Since $n \leq K$ in the first sum, the
expansion in $1/\eta$ is  correct one, and zeta function
regularization does not affect this part of the result.

We thus focus only on the second sum. To compare the zeta-function
regularized results with the integrated sum $\tilde{S}^{\rm II}$, we
first need to determine the value of the zeta function that is cut-off
$K$ dependent, and approximated by the integral as when
evaluating the sum $S^{{\rm II}}$. For this, we use simply the replacement of the
sum by an integral, as in \cite{FrolovTseytlinI}.
More precisely, we use the right colum of the following equation 
as the values of the zeta-function (taking $\alpha <1/K$)
\begin{eqnarray}
\label{cut-off-zeta}
\sum_{n=K}^{\infty} e^{- \alpha n} = {1\over \alpha} + \left(
    {1\over 2} - K   \right) + { O}(\alpha) \quad 
&\rightarrow& \quad \int_{K}^\infty dn \, e^{-\alpha n} = {1\over
    \alpha} - K + {O}(\alpha) \nonumber \\ 
\sum_{n=K}^{\infty} e^{- \alpha n} n = {1\over \alpha^2} + \left( -
    {1\over 12} + {K\over 2} - {K^2 \over 2} \right) + {
      O}(\alpha) \quad  
&\rightarrow& \quad \int_{K}^\infty dn \, e^{-\alpha n} n = {1\over
    \alpha^2} - {K^2 \over 2 } + { O}(\alpha) \nonumber \\ 
\sum_{n=K}^{\infty} e^{- \alpha n} n^2 = {2\over \alpha^3} - {1\over
  6} \left( K  - 3 K^2 + 2 K^3 \right) + { O}(\alpha) \quad
&\rightarrow& \quad \int_{K}^\infty dn \, e^{-\alpha n} n^2 = {2 \over
  \alpha^3} - {K^3 \over 3 } + { O}(\alpha) \,.\nonumber  \\
&& 
\end{eqnarray}
Note that this method was also used in
\cite{FrolovTseytlinI}. 
Comparing to the standard
Euler-Maclaurin summation formula yields that all extra tail and
boundary terms contribute subleading in $K$ and can be neglected. 
However, we will see
that this heuristic method reproduces precisely the missing terms in the
zeta-function regularization. 
We shall now compare the standard zeta-function regularized result
with the integral version zeta-function regularized expression using this
prescription. 
Let us first apply both methods to compute the sum $S^{\rm II}$ for the
folded string (\ref{ERot}) and the toy model (\ref{ToyE}).


\subsection{Folded string and toy model}

Approximating the sums (\ref{ToyE}), (\ref{ERot}) with an integral,
and subsequently expanding in $1/\eta$, we obtain, respectively
\begin{eqnarray}
\label{expann}
S_{ toy}^{\rm II}(K, \mathcal{J}) 
&=& \gamma^2 - K \gamma^2 {1\over
  {\mathcal J}} + \left( {1\over 2} K^3 \gamma^2  + {1\over 4} K
   \gamma^4 \right) {1\over {\mathcal J}^3} 
  + { O} \left( {K\over
   {\mathcal J}^5}  \right)  \nonumber\\ 
S_{ fold}^{\rm II} (K,\kappa) 
&=&  - 3 \log 2 \, \kappa^2 - 2 (\sqrt{2} -3) K \kappa - {5\over 2}
K^2  - {1\over 6} \left(\sqrt{2} - {15 \over 2}\right) {K^3 \over \kappa} + { O}
\left( {K^5 \over \kappa^3}  \right) \,. 
\end{eqnarray}
On the other hand, expanding the summands $f(n, \eta)$ as done for the
zeta-function regularization leads to
\begin{eqnarray}
\label{expd}
f_{ toy}(n, \mathcal{J}) &=&   \gamma^2 {1\over \mathcal{J}} + 
\left(-
{3\over 2} n^2 \gamma^2 - {\gamma^4 \over 4} \right) {1\over {\mathcal J}^3}
+ O\left({1\over \mathcal{J}^4}\right)  \nonumber \\
\label{expdd}
f_{fold}(n, \kappa) &=&  2 (\sqrt{2} - 3) \kappa + 5 n + \left({\sqrt{2}
  \over 2} - {15\over 4}\right) n^2 {1\over \kappa} + O\left({1\over \kappa^2}\right)  \, .
\end{eqnarray}
Comparing the expansions (\ref{expann}) with
 (\ref{expd}), we note the absence of the leading,
$1/\mathcal{J}^0$ and $\kappa^2$ terms in the expansion of the
summands. Summing up the expanded terms (\ref{expd}) from
$(K,\infty)$ and using the zeta function results (\ref{cut-off-zeta})
we obtain the same results as in (\ref{expann}) except
for the $1/\mathcal{J}^0$ and $\kappa^2$ terms, which were absent from
the beginning in the expansion.
These terms, being cut-off $K$ independent parts of the sums, can be
obtaind by setting $K=0$ in the integral. So the difference between the
two results is given by
\begin{equation}
\Delta(\eta) = \int_0^\infty f(n, \eta) \, dn \,.
\end{equation}


\subsection{The circular string in the $\sutwo$ subsector }

In this section we will consider the evaluation of the 1-loop energy
energy shift corresponding to the circular string which rotates in an
$S^3$ inside the $S^5$ with two equal spins $J_1 = J_2 =J/2$. The energy shift
takes the following form \cite{Frolov:2003tu, Frolov:2004bh, Beisert:2005mq}
\be \delta E = \delta E^{(0)} +
\sum_{n=1}^\infty \delta E^{(n)} \,, \ee 
where 
\be \ba
\label{SU2E}
\delta E^{(0)} &= 2 + \sqrt{1-{ 2 k^2\over \mathcal{J}^2 + k^2}} - 3
  \sqrt{1-{k^2 \over \mathcal{J}^2 + k^2}} \cr \delta E^{(n)} &= 2
  \sqrt{1+ {(n+ \sqrt{n^2 - 4 k^2 })^2 \over 4 (\mathcal{J}^2 + k^2
  )}} + 2 \sqrt{ 1+ {n^2 - 2 k^2 \over \mathcal{J}^2 + k^2}} + 4
  \sqrt{1 + {n^2 \over \mathcal{J}^2 + k^2}} \cr & \ \ - 8 \sqrt{1 +
  {n^2 - k^2 \over \mathcal{J}^2 + k^2}} \,.  
\ea 
\ee 
The zeta-function regularized version of the sum is derived to all
  orders in $1/\mathcal{J}$ in
  appendix A. It is hard to exactly repeat the procedure from the previous section
  for the sum (\ref{SU2E}) due to the complexity of the integral
  $\tilde{S}^{\rm II}$. So let us instead first expand the sum
  (\ref{SU2E}) in the small parameter $k$ and then repeat the
  computation from the previous section order by order in $k$.  Note
  also, that although the winding number $k$ is in principle integer
  valued, in the regime which we are interested, namely $\mathcal{J} \gg 1$,
  $n>K\gg 1$, the expansion in small $k$ is justified.

The expansion of the summand (\ref{SU2E}) is
\begin{eqnarray}\label{ElargekExp}
\delta E^{(n)} &=& - {(\mathcal{J}^2 + 2 n^2) \over \mathcal{J} n^2 (\mathcal{J}^2 + n^2)^{3 \over 2}} k^4 + {- 2 \mathcal{J}^4 - 2 \mathcal{J}^2 n^2 + n^4 \over \mathcal{J}^3 n^4 (n^2 + \mathcal{J}^2)^{3\over 2}} k^6 + O(k^8)  \,  \nonumber \\
&\equiv& \delta E_1^{(n)} k^4 + \delta E_2^{(n)} k^6 + O(k^8) \, .
\end{eqnarray}
We can now repeat the procedure from the previous section for the sums $\delta E_1$ and $\delta E_2$. 

Expansion of the first integral yields
\begin{equation}
\label{SUexpans}
\int_{K}^\infty dn\,  \delta E_1^{(n)} = - {1\over \mathcal{J} K
  \sqrt{\mathcal{J}^2 + K^2}} = - {1\over K} {1\over \mathcal{J}^2} +
    {1\over 2} K {1\over \mathcal{J}^4} - {3\over 8} K^3 {1\over
      \mathcal{J}^6} +  O\left({1\over \mathcal{J}^8}\right)\, .
\end{equation}
The integrated function thus admits an integer power expansion in
${1\over \mathcal{J}^{2n}}$, and thus is analytic in $\lambda'$.
On the other hand, the naive expansion ({\it i.e.} the expansion where we assume that  $n<\mathcal{J}$) of the integrand $\delta E_1^{(n)}$ gives
\begin{equation}
\label{f1}
\delta E_1^{(n)} = - {1\over n^2} {1\over \mathcal{J}^2}  - {1\over 2}
       {1\over \mathcal{J}^4} + {9 \over 8} n^2 {1\over
	 \mathcal{J}^6} + O\left({1\over \mathcal{J}^8}\right)\,  .
\end{equation}
As expected, these terms yield divergent sums starting from
$1/\mathcal{J}^4$, however they appear  
with powers $1/\mathcal{J}^{2k}$, {\it i.e.} the same powers of the expansion 
in (\ref{SUexpans}). Integrating the
expression (\ref{f1}) and using the integral version of the
zeta-function 
prescription (\ref{cut-off-zeta}), we reproduce
all terms in (\ref{SUexpans}).

The evaluation of the second order term $\delta E^{(n)}_2$ is
different
\begin{eqnarray}
\label{f2}
\int_{K}^\infty dn \, \delta E_2^{(n)} &=& {-2 \mathcal{J}^4 + 2
  \mathcal{J}^2 K^2 + K^3 (K - \sqrt{\mathcal{J}^2 + K^2}) \over 3
  \mathcal{J}^5 K^3 \sqrt{\mathcal{J}^2 + K^2}}  \nonumber \\ 
&=& - {2\over 3} {1\over K^3} {1\over \mathcal{J}^2} + {1\over K}
    {1\over \mathcal{J}^4} - {1\over 3} {1\over \mathcal{J}^5} -
    {1\over 4}K {1\over \mathcal{J}^6} +
  O\left({1\over \mathcal{J}^9}\right)\, . 
\end{eqnarray}
The main difference to the former case is, 
the presence of the term $1/\mathcal{J}^5$, which is non-analytic in
$\lambda'$ and
which appears as the cuf-off $K$ independent part of the integral.  On
the other hand, the naive expansion of
$\delta E_2^{(n)}$ yields
\begin{eqnarray}
\label{Eexp}
\delta E_2^{(n)} = - {1\over n^4} {1\over \mathcal{J}^2} + {1\over
  n^2}{1\over \mathcal{J}^4} + {1\over 4} {1\over \mathcal{J}^6} - {7
  \over 8} n^2 {1\over \mathcal{J}^8} +O\left({1\over
  \mathcal{J}^{10}}\right) \,,
\end{eqnarray}
where all terms are analytic in $\lambda'$. 
Since the
zeta-function prescription 
does not change the order in $1/\mathcal{J}$ in the expansion, it is thus
clear that the terms at order $1/\mathcal{J}^5$ in (\ref{f2}) can
never be reproduced by the zeta-function regularization of the expression
(\ref{Eexp}). The regular terms in (\ref{f2}) are on the other hand
easily reproduced using the
cut-off zeta-function regularization (\ref{cut-off-zeta}).  Similar
analysis for the order $k^6$ and higher, yields the discrepancy
between the zeta-function regularization and the exact string
result at the orders $1/\mathcal{J}^{2k+1}$.

A more detailed analysis of the correction terms in the
Euler-Maclaurin summation formula 
for the sums appearing in (\ref{ElargekExp}), shows that only
the coefficients of $
1/\mathcal{J}^{2k}$ are corrected, and also that all these corrections
are supressed with inverse powers of the cutoff. Thus, the approximate
integral evaluation of the coefficients of $1/\mathcal{J}^{2k+1}$ gives
the exact result\footnote{Recall that the sum $S^{I}$ in (\ref{sp})
only contributes to the even powers of $\mathcal{J}$.}.
It should be possible to resum the effect of these terms that are
missed by zeta-function regularization. 


\section*{Acknowledgments}
\label{sec:acknowledgements}

We are especially grateful to A.~Tseytlin and K.~Zarembo for useful
discussions and comments. We also thank G.~Arutyunov, N.~Beisert,
S.~Frolov, T.~Klose, K.~Peeters, J.~Plefka, M.~Staudacher and
C.~Sonnenschein for discussions. The work of S.S.-N. was partially
supported by the DFG, DAAD, and European RTN Program
MRTN-CT-2004-503369.


\setcounter{section}{0}
\appendix{Zeta-function regularization for $\sutwo$}

In this appendix we derive the all orders result that follows from
zeta-function regularization in the $\sutwo$ subsector, where 
the energy shift is (\ref{SU2E}). Up to two-loops the energy shift has
appeared recently in \cite{Minahan:2005ux}.
Evaluating the sum perturbatively in $1/\mathcal{J}^2$, \ie, $\delta E
= \sum_{i=0}^\infty \delta E_i/\mathcal{J}^{2i}$, the energy shifts at
the first three loop orders are as follows.

\noindent
$\bullet$ 1-loop:
\be
\delta E_1 = {k^2\over 2} + {1\over 2}\sum_{n}
              \left(2k^2 - n^2 + n \sqrt{n^2 - 4 k^2} \right) \,.
\ee

\noindent
$\bullet$ 2-loop:
\be
\delta E_2 = -{5 k^4 \over 8} + {1\over 8} \sum_n
              \left(-10 k^4 + n^4 - (n^2+ 2 k^2) n \sqrt{n^2 - 4
        k^2}\right)\,.
\ee
At large $n$ the sum has asymptotics $-k^4/2 + O(1/n^2)$
and thus needs to be regularized. With zeta-function
regularization $\zeta(0)= -1/2$ the energy shift is
\be
\delta E_2^{reg} = -{3 k^4 \over 8} + {1\over 8} \sum_n
                   \left(- 6 k^4 + n^4 - (n^2 + 2 k^2 )n \sqrt{n^2 - 4
           k^2}   \right) \,.
\ee

\noindent
$\bullet$ 3-loop:

\noindent
The naively expanded sum diverges as ${9k^4\over 8} n^2 + {k^6\over 4}
+ O(1/n^2)$, and needs to be regularized to give
\be
\delta E_3^{reg} = {5 k^6 \over 16} + {1\over 16} \sum_n
                   \left(10k^6 - k^4 n^2 + 2 k^2 n^4 - n^6 + (3 k^4 +
           n^4) n \sqrt{n^2 - 4 k^2}   \right) \,.
\ee

Given the relatively simple dependence on $\mathcal{J}$ of the string
frequencies, one can compute the subtraction term, necessitated by
zeta-function regularization in a closed form.
Each of the frequencies is of the type $\sqrt{1+ a/(\mathcal{J}^2 +
  k^2)}$, which has an expansion around $\mathcal{J}=\infty$.
Consider first the following term 
\be\label{FirstTerm}
\sqrt{1+ {a\over \mathcal{J}^2}} = \sum_{p=0}^\infty {1/2 \choose p}
{a^p \over \mathcal{J}^{2p} } \,.
\ee
Now, each $a$ has an expansion in $n$, and we wish to determine the
terms up to order $1/n^2$ for fixed value of $p$. Define
\be
\ba
a_1 &= {1\over 4} (n + \sqrt{n^2 - 4 k^2})^2  + k^2 \cr
a_2 &= n^2 -  k^2\,, \qquad
a_3 = n^2 + k^2\,, \qquad
a_4 = n^2  \,.
\ea
\ee
Then
\be
\ba
&\delta E^{(n)} \sqrt{1 + k^2/\mathcal{J}^2} \cr
& = \sum_{p=0}^\infty {1/2 \choose p} {1\over \mathcal{J}^{2p}}
  \left\{
     2 \left(k^2+ {(n+ \sqrt{n^2 - 4 k^2})^2 \over 4}\right)^p  + 2 (n^2 -
     k^2)^p + 4 (n^2 + k^2)^p - 8 n^{2p}
  \right\}   \cr
&= \sum_{p=0}^\infty {1/2 \choose p} {n^{2p}\over \mathcal{J}^{2p}}
  \left\{
     2 \left({1 + \sqrt{1 - 4 k^2/n^2} \over 2}\right)^p  + 2 (1 -
     k^2/n^2)^p + 4 (1 + k^2/n^2)^p - 8
  \right\}   \,.
\ea
\ee
Then invoking
\be
\left(1+ \sqrt{1 + x}\right)^p = 2^p + 2^p \sum_{q=1}^\infty
     {p-q-1\choose q-1} {p\over q} \left({x\over 4}\right)^q \,,
\ee
and the binomial theorem, we get
\be
\ba
&\delta E^{(n)} \sqrt{1+ k^2/\mathcal{J}^2 } \cr
& = \sum_{p=0}^\infty {1/2\choose p} {n^{2p}\over \mathcal{J}^{2p}}
     \left\{
     2 \sum_{q=1}^\infty {p-q-1 \choose q-1} {p\over q}
     \left(-{k^2\over n^2 }\right)^q
     + 2 \sum_{q=1}^p {p\choose q} (2+(-1)^q) \left({k^2 \over n^2}\right)^q
     \right\} \,.
\ea
\ee
Further expanding $1/\sqrt{1+ k^2/\mathcal{J}^2}$, the coefficient of
the $1/\mathcal{J}^{2p}$ term is
\be
\ba
(\delta E^{(n)})_p
& =2 \sum_{g=0}^{p}  {-1/2 \choose p-g} {1/2 \choose g}
     k^{2(p-g)} n^{2g} \times  \cr
&\qquad \qquad       \times \left\{
     \sum_{q=1}^\infty {g-q-1 \choose q-1} {g\over q}
     \left(-{k^2\over n^2 }\right)^q
     + \sum_{q=1}^g {g\choose q} (2+(-1)^q) \left({k^2 \over n^2}\right)^q
     \right\}  \,.
\ea
\ee
Again this is an unpleasant-looking hypergeometric function.
However, we only need to extract the coefficients up to the term $1/n$
of it, and in order to obtain the zero-point energy regularization, we
only have to extract the coefficient of $n^0$.
The subtraction term at order $1/\mathcal{J}^{2p}$ is
\be
\frak{S}_p=  2 \sum_{g=0}^{p} {-1/2 \choose p-g} {1/2 \choose g}
        \sum_{q=1}^g  k^{2(p+q-g)} n^{2(g-q)}(-1)^q
        \left\{
        {g-q-1 \choose q-1} {g\over q}
     +  ((-1)^q 2 + 1) {g\choose q}
     \right\} \,.
\ee
This agrees to three loops with the above explicitly obtained
expressions.
We can also determine the change to the zero-point energy, namely
\be
(\delta E^{(0)})_p^{reg} = \delta E^{(0)}_p -{1\over 2}
\left( 2k^{2p} \sum_{g=1}^p {-1/2 \choose p-g} {1/2\choose g} (1 + (-1)^g)
\right) \,.
\ee


\bibliographystyle{JHEP} \renewcommand{\refname}{Bibliography}
\addcontentsline{toc}{section}{Bibliography} 
\bibliography{main}

\end{document}